*2024-2025 CRA Quadrennial Paper*

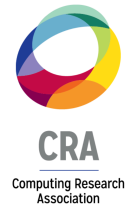
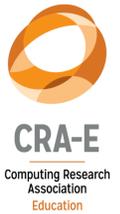

# Reversing the Computing Research Workforce Shortfall: Bolstering Domestic Student Pathways to PhDs

Susanne Hambrusch (Purdue University), Lori Pollock (University of Delaware),
Mary Hall (University of Utah), Nancy M. Amato (University of Illinois Urbana-Champaign)

**To sustain innovation and safeguard national security, the U.S. must strengthen domestic pathways to computing PhDs by engaging talented undergraduates early — before they are committed to industry — with research experiences, mentorship, and financial support for graduate studies.**

The U.S. has long relied on advancements fueled by computing research to drive innovation and sustain its global leadership in technology. However, we are facing a challenge that poses a significant threat to national security and undermines America's ability to sustain its competitive edge and innovate in key areas like artificial intelligence, cybersecurity, nanotechnology, quantum computing, and defense technologies. While the demand for computing talent is skyrocketing, the number of PhD degrees in computing among domestic students has seen only modest growth.

Between 2010 and 2023, the number of domestic computing bachelor's graduates increased by over 65,000 annually (almost a threefold increase), reflecting both a surge in student interest and computing's central role in shaping the modern world and driving innovation across industries. However, this growth has not extended to the doctoral level, where the number of domestic PhD graduates increased by only 433 students (a 0.5x increase). This loss of domestic talent from our research training pipeline threatens technological progress and limits the U.S.'s ability to cultivate the advanced expertise needed to remain competitive in critical research fields.

Strengthening domestic PhD talent is not only important for sustaining innovation but is also critical for national security. Domestic PhD graduates are better positioned to obtain security clearances, which are essential for working on sensitive government projects, especially in areas like defense, cybersecurity, and intelligence. Continuing on our current path leaves the U.S. vulnerable to disruptions that could compromise the workforce needed to maintain leadership in key research areas.

To address this challenge, we propose the **National Computing Research Workforce Fellowship**, a program that recruits promising students for graduate studies during their junior year — before their final summer internship, which often results in students accepting full-time job offers in the tech industry. This program complements and builds on existing pathways while providing a crucial new intervention, offering early exposure to research along with mentorship, motivation, and financial support for graduate studies to encourage students to pursue PhD pathways they might not have otherwise considered.

## Existing Programs that Build Interest and Open Pathways to PhD Programs

Currently, several valuable programs funded by the U.S. National Science Foundation (NSF), other government agencies, and private industry support students in pursuing research careers and graduate education. These programs focus on three key areas:

**Undergraduate Research Experiences:** Faculty-mentored projects expose students to research fields, building confidence and developing the skills needed for graduate-level research. NSF's Research Experiences for Undergraduates (REUs) fund research experiences either through REU sites or supplements to faculty-led NSF projects, while CRA's Distributed REU (DREU) program extends these opportunities by matching students with mentors at other institutions for summer research.

**Graduate Fellowships:** Programs like the NSF Graduate Research Fellowship Program (GRFP) and NSF CSGrad4US Graduate Fellowship provide critical funding for computing graduate studies, helping students focus on research without financial hardship. NSF GRFP is open to current and prospective graduate students, while NSF CSGrad4US assists individuals transitioning from the workforce back to academia. Although these fellowships offer essential support, they are limited in number and not designed to scale to the levels needed to address the growing domestic computing research workforce crisis.

**Mentoring:** The CRA's UR2PhD program provides mentorship and research opportunities for undergraduates, guiding them toward PhD programs. Similarly, NSF CSGrad4US offers mentoring and coaching on PhD applications, tailored to professionals returning from industry. Both programs help students navigate the application process and make informed decisions about pursuing a PhD.

While these programs are valuable, they often reach students already inclined toward research careers. Many other students are recruited into lucrative internships during their junior year and accept full-time jobs before considering graduate education opportunities.

## Identifying and Addressing a Key Gap

While existing initiatives show promising outcomes, a critical gap in support remains after junior year — when many students engage in summer internships that often lead to industry employment. Without strong mentoring, students may find the pathway to research and graduate school confusing and uncertain, while the route to industry appears clearer and more financially rewarding.

Our experience with thousands of students shows this moment as pivotal — with many potential PhD candidates lost to tech-sector recruitment. Our proposal for a National Computing Research Workforce Fellowship aims to address this opportune moment by providing early research opportunities, mentorship, and financial support to demonstrate the value of graduate education.

## National Computing Research Workforce Fellowship Program Structure

The National Computing Research Workforce Fellowship will offer a new pathway to computing PhDs by motivating and supporting domestic students who might not have otherwise pursued graduate studies. Building on the knowledge and resources of existing programs, the fellowship will offer:

**Extended Undergraduate Research Experiences:** National Computing Research Workforce Fellows will engage in research starting in their junior year and continue through their senior year. This extended research engagement will allow students to see the value of research careers firsthand and build relationships with their research mentors, giving them the confidence and skills needed to pursue graduate education.

**Mentoring and Career Guidance:** Fellows will receive personalized academic/career mentoring throughout the fellowship, helping them navigate the graduate application process and understand how research careers can open doors beyond immediate industry opportunities.

**Graduate Fellowship Support:** Students who complete the National Computing Research Workforce Fellowship undergraduate component and are admitted to computing PhD programs will receive fellowship funding covering tuition, stipends, and research expenses. This support ensures financial security, allowing students to focus fully on their studies without the pressure of pursuing industry opportunities.

**Graduate Internship Opportunities:** During their PhD programs, Fellows will have access to internships in government labs, federal agencies, and industry research settings, expanding their professional networks and career options in computing national service.

By integrating mentorship, research opportunities, and financial support, the National Computing Research Workforce Fellowship engages students at a critical decision point, ensuring that talented students who might have otherwise opted for industry careers are fully prepared for and inspired to pursue graduate research.

## Sustaining U.S. Leadership and National Security

The proposed National Computing Research Workforce Fellowship complements existing programs by addressing a key gap where domestic talent is often lost. By engaging students early and providing comprehensive support, it ensures that more domestic students stay on the path to advanced studies, strengthening the U.S.'s research capacity.

By inspiring students who may initially overlook the value of research careers — especially when confronted with attractive industry offers — the National Computing Research Workforce Fellowship offers a clear pathway to long-term success in research and graduate education. It also addresses the persistent, decades-long concern of increasing the number of domestic PhD graduates in computing. This strategic investment is essential to maintaining U.S. leadership in technology and safeguarding national security by building a strong, sustainable domestic research workforce capable of meeting the challenges of tomorrow.

Through this investment, the U.S. can maintain its competitive edge in technology, strengthen its domestic talent pathways, and foster a more diverse research community. The time to act is now — before more talent is lost to industry and the U.S.'s ability to innovate and lead in critical fields is compromised.

*This quadrennial paper is part of a series of papers compiled every four years by the **Computing Research Association (CRA)** and members of the computing research community to inform policymakers, community members, and the public on important research opportunities in areas of national priority. The topics chosen represent areas of mutual interest among the members spanning various subdisciplines of the computing research field. The papers attempt to portray a comprehensive picture of the computing research field detailing potential research directions, challenges, and recommendations.*

*This material is based upon work supported by the U.S. National Science Foundation (NSF) under Grant No. 2123180, 2231962, 2313998, 2417847, and 2216270. Any opinions, findings, and conclusions or recommendations expressed in this material are those of the authors and do not necessarily reflect the views of NSF.*